\documentstyle[12pt]{article}
\newcommand{\beq}{\begin{equation}}
\newcommand{\eeq}{\end{equation}}
\newcommand{\bra}{\begin{array}}
\newcommand{\era}{\end{array}}
\newcommand{\be}{\beta}
\newcommand{\al}{\alpha}

\newcommand{\de}{\delta}

\begin{document}
\begin{center}
{\Large\bf  On the $C_{\lambda}$-extended $w_{\infty}$-symmetry }
\vskip1.4truecm {\bf J. Douari}\footnote{ Permanent Address:
Universit\'e Mohammed V, Facult\'e des Sciences, D\'epartement de
Physique, LPT, Av. Ibn Batouta, B.P. 1014, Agdal, Rabat, Morocco}
\vskip.2truecm Institut des Hautes Etudes Scientifiques, Le
Bois-Marie, 35, route de Chartres, F-91440
Bures-sur-Yvette, France.\\
\vskip1 truecm {\bf E. H. El Kinani}\footnote{ Junior Associate at
The Abdus Salam ICTP, Trieste, Italy. \\Permanent Address:
Universit\'e Moulay Ismail, Facult\'e des Sciences et Techniques,
D\'epartement de Math\'ematiques, Laboratoire de Physique
Math\'ematique, Boutalamine B.P.509, Errachidia, Morocco.}\\
The Abdus Salam International Center for Theoretical Physics,
Strada Costiera 11,
 34014, Trieste, Italy.
\end{center}

\section*{Abstract}
\hspace{.3in}Starting from the $C_{\lambda}$-extended oscillator
algebras, we obtain a new deformed $w_{\infty}$-algebra. More
precisely, we show that the $C_{\lambda}$-extended
$w_{\infty}$-algebra generators may be expressed via the
annihilation and creation operators of the $C_{\lambda}$-extended
oscillator algebras $a$ and $a^{\dagger}$ as an
infinite-dimensional extension of the realization of $sp(2)$
algebra.

\section{Introduction}
\hspace{.3in}Over the past decade, much attention has been paid to
the oscillator algebras in connection with the important role
investigated in many physical systems, and they provide an useful
tool in the theory of Lie algebra representations. On the other
hand, their deformations and extensions have given new
mathematical objects called Quantum Groups or Quantum Algebras
\cite{1} which have found applications in various physical
problems, such as the description of the fractional statistics
\cite{2}, the bosonization of supersymmetric quantum mechanics
\cite{3,4} and in the  $n$-particle integrable systems \cite{4,5}.
Among these various deformations and extensions, we mention the
generalized deformed oscillator algebras (GDOA's) \cite{6} and the
$G$-extended oscillator algebras \cite{7} where $G$ is some finite
group, e.g in the case of Calogero-model
$C_{\lambda}=\mathcal{Z}_{\lambda}$ is the cyclic group of order
$\lambda$. This latter, have recently proved very useful in the
context of supersymmetric quantum mechanics and some of its
variants \cite{7,8}, and also in the study of coherent states in
nonlinear quantum optics \cite{2}.\\

In this paper, we recall some facts concerning the
$C_{\lambda}$-extended oscillator algebras which were introduced
as a generalization of Calogero-Vasiliev algebras \cite{4,9}. In
the second section  we give some of their mathematical properties,
and we extend the commutation relation between annihilation and
creation operators $a$ and $a^{\dagger}$ to any indices $n$ and
$m$, as we will see for the positive indices it is
straightforward. However, for the formal negative powers one
consider the Bargmann representation \cite{9}, indeed in such
representation the generators $a$ and $a^{\dagger}$ are identified
with differential operators acting on the Bargmann space. Section
three is devoted to the construction of $w_{\infty}$-algebra
generators in terms of the pair $a$ and $a^{\dagger}$ of the
$C_{\lambda}$-extended oscillator algebras, we begin by the case
of the Virasoro generators. We enclose by a conclusion and
remarks.
\section {Review on $C_{\lambda}$-extended oscillator algebras}
\hspace{.3in}As a generalization of the Calogero-Vasiliev
algebras, the $C_{\lambda}$-extended oscillator algebras (also
called generalized deformed oscillator algebras (GDOA's)), denoted
$A^{\lambda}$, $\lambda=2,3,...,$ are defined by \beq \bra{lrlr}
[N,a^{\dagger}]=a^{\dagger},& [N,P_{\mu}]=0,\\ \\

[a,a^{\dagger} ]= I+\sum\limits_{\mu=0}^{\lambda-1}
\alpha_{\mu}P_{\mu}, & a^{\dagger}P_{\mu}=P_{\mu +1}a^{\dagger},
\era \eeq together with their hermitian conjugates, and \beq
\bra{lrlr}
P_{\mu}=\frac{1}{\lambda}\sum\limits_{\nu=0}^{\lambda-1}
e^{\frac{2\pi i\nu(N-\mu)}{\lambda}},&
\sum\limits_{\mu=0}^{\lambda-1}P_{\mu}=1&
P_{\mu}P_{\nu}=\de_{\mu,\nu}P_{\nu}\\ \\
\sum\limits_{\mu=0}^{\lambda-1}\alpha_{\mu}=0,&
\sum\limits_{\nu=0}^{\mu-1}\alpha_{\nu}>-1,& \mu=1,...,\lambda-1.
\era, \eeq  where $\alpha_{\mu}\in{\bf R\rm}$, $N$ is the number
operator and $P_{\mu}$ are the projection operators on subspaces
$F_{\mu}=\{\vert k\lambda -\mu\rangle \vert k=0,1,2,...\}$ of the
Fock space $F$ which is portioning into $\lambda$ subspaces.\\

The operators $a$ and $a^{\dagger}$ are defined by \beq \bra{lrlr}
a^{\dagger}a=F(N),& aa^{\dagger}=F(N+1), \era \eeq where
$F(N)=N+\sum\limits_{\mu=0}^{\lambda-1}\be_{\mu}P_{\mu}$,
$\be_{\mu}=\sum\limits_{\nu=0}^{\mu-1}\al_{\nu}$, which is a
fundamental concept of deformed oscillators. Let's  denote the
basis states of subspaces $F_{\mu}$ by $\vert n\rangle=\vert
k\lambda +\mu\rangle\simeq (a^{\dagger})^n\vert 0\rangle$ where
$a\vert 0\rangle=0$, $\vert 0\rangle$ is the vacuum state. The
operators $a$, $a^{\dagger}$  and $N$ act on $F_{\mu}$ as follows
\beq \bra{rcl} N\vert n\rangle=n\vert n\rangle,& a^{\dagger}\vert
n\rangle=\sqrt{F(N+1)}\vert n+1\rangle,& a\vert
n\rangle=\sqrt{F(N)}\vert n-1\rangle. \era \eeq According to these
relations, $a$ and $a^{\dagger}$ are the
annihilation and the creation operators respectively.\\

Particularly, if $\lambda=2$, we have two projection operators
$P_{0}=\frac{1}{2}(I+(-1)^{N})$ and
$P_{1}=\frac{1}{2}(I-(-1)^{N})$ on the even and odd subspaces of
the Fock space $F$, and the relations of (1) are restricted to
\beq \bra{rcl} [N,a^{\dagger}]=a^{\dagger},&
[a,a^{\dagger}]=I+\kappa K,& \{K,a^{\dagger}\}=0, \era \eeq with
their hermitian conjugates, where $K=(-1)^{N}$ is the Klein
operator and $\kappa$ is a real parameter. These relations define
the so-called Calogero-Vasiliev algebra.\\

The $C_{\lambda}$-extended oscillator algebras are seeing as
deformation of $G$-extended oscillator algebras, where $G$ is some
finite group, appeared in connection with $n$-particle integrable
models. In the former case, $G$ is the symmetric group $S_n$. So,
for two particles $S_2$, can be realized in terms of $K$ and
$S_2$-extended oscillator algebra becomes a generalized deformed
oscillator algebra (GDOA) also known as the Calogero-Vasiliev or
modified oscillator algebra. In the $C_{\lambda}$-extended
oscillator algebras, $G\equiv C_{\lambda}$ is the cyclic group of
order $\lambda$, $C_{\lambda}=\{1,K,...,K^{\lambda-1}\}$. So,
these algebras have a rich structure since they depend upon
$\lambda$ independent real parameters, $\alpha_0 ,\alpha_1
,...,\alpha_{\lambda-1}$.\\

Thus, these algebras can be also rewritten in terms of Klein
operator, and instead of (1) we have \beq \bra{rcl}
[N,a^{\dagger}]=a^{\dagger},& [N,K]=0,& K^{\lambda}=I, \era \eeq

\beq \bra{rl} [a,a^{\dagger}]=I+\sum\limits_{r=1}^{\lambda-1}
\kappa_r K^r, & a^{\dagger}K=e^{\frac{-2\pi
i}{\lambda}}Ka^{\dagger}, \era \eeq together with their hermitian
conjugates, where
$\kappa_{\mu}\in{\bf C\rm}$ and $\kappa_{\mu}^* =\kappa_{\lambda -\mu}$.\\

To see the equivalence between (1) and (6-7), we note that
$C_{\lambda}$ has $\lambda$ inequivalent one-dimensional unitary
irreducible matrix representations $\Gamma^{\mu}$, $\mu
=0,1,...,\lambda-1$. The projection operator on the carrier space
of $\Gamma^{\mu}$ can be written as \beq
P_{\mu}=\frac{1}{\lambda}\sum\limits_{\nu=0}^{\lambda-1}(\Gamma^{\mu}(K^{\nu}))^*
K^{\nu}=\frac{1}{\lambda}\sum\limits_{\nu=0}^{\lambda-1}e^{-2i\pi\mu\nu}K^{\nu},
\eeq and $\alpha_{\mu}$ can be  given in terms of $\kappa_{\mu}$
as follows \beq \bra{rl}
\alpha_{\mu}=\sum\limits_{\nu=0}^{\lambda-1}e^{\frac{-2i\pi\mu\nu}{\lambda}}\kappa_{\nu},&
\mu=0,1,...,\lambda-1. \era \eeq

The $C_{\lambda}$-extended oscillator Hamiltonian defined in the
bosonic Fock space representation as usual by \beq H_0
=\frac{1}{2}\{a,a^{\dagger}\}. \eeq Owing to the relations of
(6-7) the Hamiltonian can be rewritten in terms of the projection
operators as \beq \bra{rcl} H_0
=N+\frac{1}{2}I+\sum\limits_{\mu=0}^{\lambda-1}\gamma_{\mu}P_{\mu},&
\gamma_{\mu}\equiv\frac{1}{2}(\be_{\mu}+\be_{\mu+1}),&
\be_{\mu}=\sum\limits_{\nu=0}^{\mu-1}\al_{\nu}. \era \eeq Its
eigenstates are $\vert n\rangle=\vert k\lambda+\mu\rangle$ and
their eigenvalues are given by \beq \bra{rcl}
E_{k\lambda+\mu}=k\lambda+\mu+\gamma_{\mu}+\frac{1}{2},&
k=0,1,...,& \mu=0,1,...,\lambda -1. \era \eeq

In each subspace $F_{\mu}$, the spectrum of $H_0$ is harmonic, but
the $\lambda$ infinite sets of equally spaced energy levels,
corresponding to $\mu=0,1,...,\lambda -1$, can be shifted with
respect to each other by some amounts depending upon the
parameters $\al_0 ,\al_1 ,...,\al_{\lambda -1}$.\\

In what follows, we will extend the commutation relations between
$a^{\dagger}$ and $a$. First for the positive indices the
extension is straightforward then from the following commutation
relation between the generators $a^{\dagger}$ , $a$ and $K$.

\begin{equation}
\left[ a, a^{\dagger} \right] = I+
\sum_{r=1}^{\lambda-1}{\kappa}_{r}{K}^{r} \:\:\ , \:\:\
a^{\dagger}K= e^{-2 \pi i / \lambda} K a^{\dagger},
\end{equation}
one obtain easily after algebraic manipulation the following
relations \cite{17}

\begin{equation}
\left[ a, (a^{\dagger})^{m} \right] =( m +
\sum_{r=1}^{\lambda-1}f_{r}{\kappa }_{r}
{K}^{r})(a^{\dagger})^{m-1},
\end{equation}
where the functions $f_{r}$ are given by \beq \bra{lrlr}
 f_{r}=\sum_{p=0}^{m-1}  e^{-i\frac{2r\pi}{\lambda}p},& r=2,3,...,\lambda-1,
\\
f_{1}=1.& \era \eeq  The negative values $m<-1$ are described as
monomials of $a^{\dagger}$ with negative powers, which act as
differential operators on the Bargmann representation \cite{9}.
For any order $n$ of $a$ one obtain,

\begin{equation}
[a^n
,(a^{\dagger})^m]=\sum_{\alpha=0}^{n-1}\sum_{l=0}^{\alpha}\big (
m+ F x^{\alpha}  \big ) \beta_l (a^{\dagger})^{m-l-1}a^{n-l-1},
\end{equation}
where
\begin{equation}
\begin{array}{rl}
F= \sum\limits_{r=1}^{\lambda-1} f_r \kappa_r K^r ,&
x=e^{\frac{-2\pi i}{\lambda}},
\end{array}
\end{equation}
and such that

\begin{equation}
\begin{array}{ll}
a^n (a^{\dagger})^{m-1}&= \beta_{n}
(a^{\dagger})^{m-n-1}+\beta_{n-1} (a^{\dagger})^{m-n}a+
\beta_{n-2} (a^{\dagger})^{m-n+1}a^2 +...+\\ \\
&\beta_{3} (a^{\dagger})^{m-4}a^{n-3}+\beta_{2}
(a^{\dagger})^{m-3}a^{n-2}+ \beta_{1}
(a^{\dagger})^{m-2}a^{n-1}+\beta_{0} (a^{\dagger})^{m-1}a^{n}.
\end{array}
\end{equation}
Owing to the relations (13) and (14) the parameters $\beta_{i}$,
$i=0,1,2,...,n-k,...,n$, are expressed by the following equalities

\begin{equation}
\begin{array}{llllll}
\beta_{n}  &= \prod\limits_{i=1}^{n} \Big ( (m-i)+ F x^{n-i} \Big ) .\\ \\

\beta_{n-1}&= \prod\limits_{i=1}^{n-1} \Big ( (m-i)+ F x^{n-i}\Big
) +\sum\limits_{\mu=2}^{n-2}\prod\limits_{i=1}^{n-\mu} \Big (
(m-i)+F x^{n-i} \Big )
\prod\limits_{i=n-(\mu-1)}^{n-1} \Big ( (m-i)+ F x^{\mu-i-1} \Big )+\\ \\
&\Big ( 2(m-1)+\sum\limits_{j=n-2}^{n-1} F x^j \Big )
\prod\limits_{i=2}^{n-1} \Big ((m-i)+ F x^{n-i-1}\Big ) .\\
& .\\
& .\\
& .\\
\beta_{n-k}&= \prod\limits_{i=1}^{n-k} \Big ( (m-i)+ F x^{n-i}\Big
) + \prod\limits_{i=1}^{n-k-1} \Big ( (m-i)+F x^{n-i} \Big )
\Big ( k(m-(n-k))+\sum\limits_{j=0}^{k-1} F x^j \Big )+\\  \\
&\sum\limits_{\al=k+2}^{n-2}\prod\limits_{i=1}^{n-\al} \Big (
(m-i)+F x^{n-i} \Big )
\Big [ \prod\limits_{i=n-\al+1)}^{n-k} \Big ( (m-i)+ F x^{n-i-1} \Big )+\\ \\
&\sum\limits_{\mu=n-\al+1}^{n-k-1}\prod\limits_{i=n-\al-1}^{n-\mu}
\Big ( (m-i)+F x^{n-i-1} \Big )
\prod\limits_{i=n-\mu+1}^{n-k} \Big ( (m-i)+ F x^{\mu-i-2} \Big )+\\ \\

&\sum\limits_{l=2}^{k}\Big (
l(m-(n-\al+1))+\sum\limits_{j=\al-l-1}^{n-4} F x^j \Big )
\prod\limits_{i=n-\al+2}^{n-l} \Big ((m-i)+ F x^{n-i-l}\Big )\Big ]+\\ \\

&\sum\limits_{l=2}^{k}\Big ( l(m-1)+\sum\limits_{j=n-2}^{n-1} F
x^j \Big )
\Big [ \prod\limits_{i=2)}^{n-k} \Big ( (m-i)+ F x^{n-i-1} \Big )+\\ \\

&\sum\limits_{\mu=k+1}^{n-4}\prod\limits_{i=2}^{n-\mu} \Big (
(m-i)+F x^{n-i-1} \Big )
\prod\limits_{i=n-\mu+1}^{n-k-2} \Big ( (m-i)+ F x^{\mu-i-2} \Big )+\\ \\

&\Big ( (m-2)+ F x^{n-3}\Big ) \Big (
(k+2-l)(m-3)+\sum\limits_{j=n-5}^{n-4} F x^j \Big )
\prod\limits_{i=n-k-1}^{n-k} \Big ((m-i)+ F x^{n-i-k}\Big )\Big ] +\\ \\

&\Big ( (k+1)(m-1)+\sum\limits_{j=2}^{n-1} F x^j \Big )
\prod\limits_{i=2}^{n-2} \Big ((m-i)+ F x^{n-i-2}\Big ) ,\:\:\:\:\  k=2,...,n-4\\
& .\\
& .\\
& .\\
\beta_{3}&=\prod\limits _{i=1}^{3} \Big ( (m-i)+F x^{n-i} \Big )
+\prod\limits_{i=1}^{2} \Big ( (m-i)+F x^{n-i} \Big )
\Big ( (n-3)(m-3)+\sum\limits_{j=0}^{n-4} F x^j \Big )+\\ \\
&\sum\limits_{\mu=2}^{n-2} \Big (
\mu(m-1)+\sum\limits_{j=n-\nu}^{n-1} F x^j \Big ) \Big ( (m-2)+F
x^{n-\mu-1}\Big )\Big ( (n-\mu-1)(m-3)+\sum\limits_{j=0}^{n-\mu-2}
F
x^{j} \Big )\\ \\

\beta_{2}&= \prod \limits_{i=1}^{2} \Big ( (m-i)+F x^{n-i} \Big
)+\sum\limits_{\mu=2}^{n-1} \Big (
\mu(m-1)+\sum\limits_{j=n-\mu}^{n-1} F x^j
\Big )\Big ( (m-2)+Fx^{n-\mu-1} \Big )\\ \\

\beta_{1}&= \Big ( n(m-1)+\sum\limits_{j=0}^{n-1} F x^j \Big )\\
\beta_{0}&=1\\

\end{array}
\end{equation}
\section {Modified $w_{\infty}$-symmetry}
\hspace{.3in}In this section, we introduce the new modified
$w_{\infty}$-symmetry using the $C_\lambda$-extended oscillator
generators. Here we will adopt the approach for deformed or
undeformed case where the elements of the $w_{\infty}$-algebra
$\omega^{s}_{m}$ are expressed in terms of the pair $a^{\dagger}$
and $a$ of quantum oscillator with or without deformation. Before
going on, let us first restrict ourselves to the case of classical
Virasoro algebra generated by the operators $\omega^{2}_{m} =
\ell_{m}$
\subsection  {Classical conformal symmetry and its representations}
\hspace{.3in}For reader convenience we remind some information
about the conformal symmetry termed also Virasoro symmetry which
was first introduced in the context of string theories and it is
relevant to any theory in (2+1)-dimensional space-time which
possesses conformal invariance. The conformal symmetry consists of
all general transformations
\begin{equation}
z\to z+\epsilon(z), \:\:\:\:\ \bar{z} \to \bar{z} +
\bar{\epsilon}(\bar{z})
\end{equation}
where $\epsilon(z)$ and $(\bar{\epsilon}(\bar{z}))$ are an
infinitesimal analytical (anti-analytical) functions. It can be
represented as an infinite Laurent series
\begin{equation}
\epsilon(z)=\sum_{n}\epsilon_{n}z^{n+1},  \:\:\:\:\ n \in
\mathcal{Z},
\end{equation}
and an analogous formula holds for $\bar{\epsilon}(\bar{z})$.\\

These mapping are generated by the differential operators
\begin{equation}
\ell_{m} = -z^{m+1}\partial_z  \:\:\  \mbox {and} \:\:\
\bar{\ell}_{m} = - \bar{z}^{m+1}\partial_{\bar{z}}.
\end{equation}
The operators $\ell_{m}$ satisfy the commutator relations
\begin{equation}
[\ \ell_{m}, \ell_n] = (m-n)\ell_{m+n},
\end{equation}
and an analogous formula holds for the $\bar{\ell}_{m}$. Hence,
this means that both the $\ell_{m}$ and the $\bar{\ell}_{m}$ span
an infinite-dimensional Lie algebra; moreover, these two algebras
are combined as a direct sum, $[\ell_{m},\bar{\ell}_{m}]=0$. The
algebra defined by (23) is known as the classical conformal or
Virasoro algebra. We shall denote that the above algebra admits a
unique $1$-dimentional central extension
\begin{equation}
{\mathcal{L}}_{c}= {\mathcal{L}} \oplus {\mathcal{C}}c,
\end{equation}
with the following commutation relations
\begin{equation}
[\ \ell_{m}, c] = 0, \:\:\:\:\:\  [\ell_{m}, \ell_n] =
(m-n)\ell_{m+n} +c {\frac{(m^{3}-m)}{12}}\delta_{m+n,0},
\end{equation}
where the value of the central charge $c$ is the parameter of the
theory in the quantum field theory context (e.g $c=1$ for the free
boson). The unitary representation of the above algebra are well
know (see e.g \cite{10}). In what follows, we give some
representations
discussed in the literature,\\

(i)  the oscillatory representation of the elements $\ell_{m}$
$$\ell_{m} =(a^{\dagger})^{m+1}a,$$

(ii) the Sugawara construction, where the generators $\ell_{m}$
are constructed in terms of infinite set of free boson operators
($b_{m},b_{m}^{\dagger}=b_{-m}, m=1,2,...$) as

$$\ell_{m}={\frac{1}{2}}\sum_{j\in Z}:b_{m-j}b_{j}:,$$ where the colons indicate
normal ordering,
\\

(iii)  the infinite matrix realization $$ \ell_{m}=\sum_{j\in
Z}E_{m,m-j}, $$ where $E_{k,j} $ are the infinite dimensional
matrix with value $1$ at the $(k,j)$
entry and $0$ at all other entries \cite{10}.\\

Here we consider the simplest one which is given by oscillatory
representation. In fact using the commutation relations of the
quantum mechanics, we obtain the following commutator
\begin{equation}
[a,(a^{\dagger})^{m}]= m (a^{\dagger})^{m-1}.
\end{equation}
From this commutator, one see easily that the Virasoro operators
for the unidimensional harmonic oscillators can be defined by
\begin{equation}
\ell_{m} =(a^{\dagger})^{m+1}a,
\end{equation}
which satisfy the following centerless Virasoro algebra
\begin{equation}
[\ \ell_{m}, \ell_n] = (m-n)\ell_{m+n}.
\end{equation}
\subsection {$C_\lambda$-deformed Virasoro algebra}
\hspace{.3in}First, we recall that  the deformation of the
conformal  algebra was first introduced by Curtright and Zachos
\cite{11} and investigated on many occasions by many authors
\cite{12,13,14}, and defined by the following $q$-commutation
relation
\begin{equation}
[\ \ell_{m}, \ell_n]_{q} =q^{m-n} \ell_{m} \ell_n-q^{n-m}\ell_n
\ell_{m}= \frac {(q^{m-n}-q^{n-m})}{q - q^{-1}} \ell_{m+n}.
\end{equation}

Turn now to the construction of the $C_\lambda$-deformed Virasoro
algebra. To do this, we will adopt the approach for undeformed
case \cite{15} where the generators $\ell_{m}$ are constructed
from one undeformed oscillator pair $a^{\dagger}$ and $a$ eq (27)
as the infinite-dimensional extension of the following realization
of $sp(2) \sim o(2,1)$

\begin{equation}
\ell_{-1}=a, \:\:\:\ \ell_0=(a^{\dagger})a, \:\:\:\
\ell_1=(a^{\dagger})^{2}a.
\end{equation}

The extension to positive indices $m$ is straightforward

\begin{equation}
\ell_{m}(\kappa)=(a^{\dagger})^{m+1}a,
\end{equation}
and for the negative values ($m <-1$, which makes this
construction formal) the generators $\ell_{m}(\kappa)$ are
described by the nonanalytic dependence (monomials of
$a^{\dagger}$ with negative powers), which act as the differential
operators in the Bargmann
representation \cite{9}.\\

Then from the previous equations (14, 15), one gets the
commutation relations for the generators $ \ell_{m} (\kappa)$ and
the Klein operator $K$ \cite{17}
\beq \left[\ell_{m}(\kappa),
\ell_{n} (\kappa)\right ] =(m-n) \ell_{m+n} (\kappa)+
\sum_{r=1}^{\lambda-1}(e^{(n+1)r(2 \pi i / \lambda)}-e^{(m+1)r(2
\pi i / \lambda)}){\kappa}_{r} {K}^{r} \ell_{m+n} (\kappa) \eeq
\beq  \left[\ell_{m} (\kappa), K \right] = g^{m}(\lambda)\ell_{m}
(\kappa)K, \eeq where the functions $g^{m}(\lambda)$ are given by
\begin{equation}
g^{m}(\lambda) =(1- e^ \frac{2 \pi i(m+1)}{\lambda}).
\end{equation}
From the above expression of the function $g^{m}(\lambda)$, one
note that the generators $\ell_{m}(\kappa)$ commutes with the
Klein-operator only in the case where $(1- e^ \frac{2 \pi
i(m+1)}{\lambda}=0)$ that is when $m+1=k\lambda, \:\:\ k \in
\mathcal{Z}$. Note also that the ordinary Virasoro algebra is
obtained for ${\kappa}_{r} \to 0$. However, when ${\kappa}_{r} \ne
0$ is something new. In the case of the Calogero-Vasiliev case,
$\lambda=2$ $(r=1)$, the relation (14) becomes

\begin{equation}
\left[ a, (a^{\dagger})^{m} \right] = (m + (1-(-1)^{m}){\kappa}
_{1} {K})(a^{\dagger})^{m-1},
\end{equation}

In what follows, we investigate the possible cases of the indices
$m$: (I) $m=2k$ $(k \in \mathcal{Z})$ even and (II) $ m=2k+1 $
odd. In (I) the above commutation relation becomes
\begin{equation}
\left[ a, (a^{\dagger})^{2k} \right] = 2k (a^{\dagger})^{2k-1},
\end{equation}
and the obtained $C_{2}$-Virosoro algebra is given by
\begin{equation}
\left[ \ell_{2k} (\kappa), \ell_{2l} (\kappa) \right] =
(2k-2l)\ell_{2k+2l}, \:\:\:\:\ [\ell_{2k}, K]=2\ell_{2k}K.
\end{equation}
In the case (II), we have

\begin{equation}
\left[ a, (a^{\dagger})^{2k+1} \right] = (2k+1 +2{\kappa} _{1}
{K}) (a^{\dagger})^{2k},
\end{equation}
we obtain the following algebra
\begin{equation} \left[ \ell_{2k+1} (\kappa),
\ell_{2l+1} (\kappa) \right] = (2k-2l)\ell_{2k+2l+2}, \:\:\:\:\
[\ell_{2k+1}, K]=0.
\end{equation}

Then, the conformal algebra is not changed when the two indices
$(m, n)$ are both even or odd, however when one indices is even
and the other is odd the obtained Virasoro algebra is changed, we
have \beq \left[\ell_{2n} (\kappa), \ell_{2m+1} (\kappa)\right ]
=2(m-n) \ell_{2m+2n+1} (\kappa)+ (1-2\kappa_{1}K)\ell_{2m+2n+1},
\eeq \beq \left[\ell_{2n} (\kappa), K \right] = 2\ell_{2n}
(\kappa) K \eeq \beq \left[\ell_{2m+1} (\kappa), K \right] = 0
\eeq

\subsection {$C_\lambda$-deformed $w_{\infty}$-algebra}

\hspace{.3in} In this section, we will extended the construction
done in the previous section on $w_{\infty}$-algebra which is
seeing as a generalization of Virasoro algebra. This algebra is
interesting in the conformal field theory point of view because it
contains the centerless Virasoro algebra as subalgebra and thus
may be viewed as an extended conformal symmetry with an infinite
set of additional symmetries. The $w_{\infty}$-algebra contains
generators  $ \omega^{s}_{m}$ of spin $2,3,4,...$ The defining
commutation relations are
\begin{equation}
[\omega^{s}_{m} ,\omega^{t}_{n}]=\Big ( (t-1)m-(s-1)n\Big )
\omega^{s+t-2}_{m+n}.
\end{equation}
The generators $ \omega^{2}_{m}$  generate a classical Virasoro
subalgebra. The conformal properties of these generators can be
viewed from the commutation relation
\begin{equation}
[\omega^{2}_{m} ,\omega^{t}_{n}]=\Big ( (t-1)m-n\Big )
\omega^{t}_{m+n}.
\end{equation}
which implies that $ \omega^{t}_{m}$ can be regarded as the
algebra Fourier modes of primary fields with the conform  spin
$t$. On the other hand $w_{\infty}$-algebra can has a natural
geometric interpretation as area preserving diffeomorphisms of
$2$-manifolds e.g on the two-dimensional cylinder  ${\mathcal{R}}
\times S^{1}$ with canonical coordinates $x$ and $y$ it can be
represented by Poisson bracket of the smooth function $
\omega^{s}_{m}=e^{mx}y^{s-1}$. However, the standard central
extension of Virasoro algebra can not extended to the full algebra
$w_{\infty}$, which should therefore be viewed as a classical
$W$-algebra. The $W_{\infty}$-algebra, which was first introduced
in refs \cite{18}, is a deformation of $w_{\infty}$ which is such
that the standard central term in the Virasoro subalgebra can be
extended to the whole algebra. $W_{\infty}$-algebra can thus be
viewed as the quantum version of $w_{\infty}$. This has been
nicely illustrated in the context of $W$ gravity. Denote the
generators of $W_{\infty}$-algebra of spin $s$ by $V^{i}_{m}$,
where $s=i+2$, and the index $i$ ranges from $0$ to ${\infty}$.
The defining commutation relations of $W_{\infty}$-algebra  can be
written as
\begin{equation}
[V^{i}_{m} ,V^{j}_{n}]= \sum_{l>0}g_{2l}^{ij}(m,n)V_{m+n}^{i+j-2l}
+c_{i}(m)\delta^{ij}\delta_{m+n}.
\end{equation}

The structure constant $g_{2l}^{ij}(m,n)$ and the central terms
$c_{i}(m)$ are completely fixed by the Jacobi identities and take
the following form
\begin{equation}
c_{i}(m)=m(m^{2}-1)(m^{2}-4)...(m^{2}-(i+1)^{2})c_{i},
\end{equation}
where the central charges $c_{i}$ are given  by
\begin{equation}
c_{i}=\frac{2^{2i-3}i!(i+1)!}{(2i+1)!!(2i+3)!!}.
\end{equation}
The structure constant  $g_{2l}^{ij}(m,n)$ are given by

\begin{equation}
g_{2l}^{ij}(m,n)=\frac{1}{2(l+1)}\phi_{l}^{ij}N_{l}^{ij}(m,n),
\end{equation}
where the $N_{l}^{ij}(m,n)$ are expressed as
\begin{equation}
N_{l}^{ij}(m,n)=\sum_{k=0}^{l+1}(-1)^{k}(_{k}^{l+1})[i+1+m]_{(l+1-k)}[i+1+m]_{k}[j+1+n]_{k}[j+1-n]_{(l+1-k)}
\end{equation}
where $[x]_{n}=\frac{\Gamma(x+1)}{\Gamma(x+1-n)}$. Finally, the
$\phi_{l}^{ij}$ are given by
\begin{equation}
\phi_{l}^{ij}=\sum_{k>0}\frac{(\frac{-1}{2})_{k}(\frac{3}{2})_{k}(\frac{-1l}{2}
-\frac{-1}{2})_{k}
(\frac{-1l}{2})_{k}}{k!(-i-\frac{-1}{2})_{k}(-j-\frac{-1}{2})_{k}(i+j-l+\frac{5}{2})_{k}},
\end{equation}
with the notation $(x)_{n}=\frac{\Gamma(x+n)}{\Gamma(x)}=x(x+1)...(x+n-1).$\\

Now, turn the realization of $w_{\infty}$-algebra generators in
terms of the pair $a^{\dagger}$ and $a$ of the
$C_\lambda$-extended oscillator algebras by extending the
realization (31). To begin, let'us define the following operators
\begin{equation}
\omega^{s}_{m}=(a^{\dagger})^{s}a^{m}
\end{equation}
By using the previous tools (16-19) the commutation relations of
the operators $\omega^{s}_{m}$ are

\beq [ \omega^{s}_{m} ,\omega^{t}_{n} ]=(\Xi^{(s,m)} -
\Xi^{(t,n)}) \omega^{s+t-(l+1)}_{m+n-(l+1)}, \eeq where \beq
\Xi^{(s,m)}=m(t+F x^{s}){\beta}_{0}
^{[s]}+(m-1)(t+Fx^{1+s}){\beta}_{1}^{[s]}+( m-2)(t+Fx^{2+s})
{\beta}_{2} ^{[s]}+...+(t+Fx^{(m-1)+s}){\beta}_{m-1}^{[s]},\eeq
and
 \beq
\Xi^{(t,n)}= n(s+Fx^{t}){\beta}_{0} ^{[t]}+(n-1)(s+Fx^{1+t})
{\beta}_{1} ^{[t]}+(n-2)(s+Fx^{2+t}){\beta}_{2} ^{[t]}+...
+(s+Fx^{(n-1)+t}){\beta}_{n-1}^{[t]},\eeq where the notation
$\beta_l ^{[n]} $ means that each $x^j$ in $\beta_l $ becomes
$x^{jn}$ and $\beta_l $ are given by eq (19). Owing to the
commutation relations of (13) and the definition of
$\omega^{s}_{m}$, we obtain the following commutation relations
between the $K$ (Klein operator) and the generators
$\omega^{s}_{m}$. \beq \bra{rl} [\omega^{s}_{m} ,K]=(x^{s+m }-1)K
\omega^{s}_{m}. \era \eeq We note that only in the case where
$s+m= \lambda$, the generators
$\omega^{s}_{m}$ commute with the Klein operator.\\

In what follows, we will examine some particular cases obtained as
limits. First, setting  $m=1$ and $ s\to s+1$, we recover the
 $C_{\lambda}$-extended Virasoro algebra (32). Second, the $C_{\lambda}$-extended
 $sp(2)$ is generated by $\omega^{0}_{1}, \omega^{2}_{1}$ and $\omega^{1}_{1}$ which
satisfy \beq [\omega^{0}_{1},\omega^{1}_{1}
]\equiv\sum\limits_{\alpha ,l}\Phi_{\alpha
,l}^{[0,1,1,1]}\omega^{-l}_{1-l} \eeq
\beq[\omega^{2}_{1},\omega^{1}_{1} ]\equiv\sum\limits_{\alpha
,l}\Phi_{\alpha ,l} ^{[2,1,1,1]}\omega^{2-l}_{1-l}  \eeq \beq
[\omega^{2}_{1},\omega^{0}_{1} ]\equiv\sum\limits_{\alpha
,l}\Phi_{\alpha ,l} ^{[2,1,0,1]}\omega^{1-l}_{1-l}, \eeq where

\beq \sum\limits_{\alpha,l}\Phi_{\alpha ,l}
^{[0,1,1,1]}\omega^{-l}_{1-l}=[\sum\limits_{r=1}^{\lambda-1}\kappa_r
K^r (1-x)+1]\omega^{0}_{1} \eeq \beq
\sum\limits_{\alpha,l}\Phi_{\alpha ,l}
^{[2,1,1,1]}\omega^{2-l}_{1-l}=[\sum\limits_{r=1}^{\lambda-1}\kappa_r
K^r (1+x^r ) x(x-1)-1]\omega^{2}_{1} \eeq \beq
\sum\limits_{\alpha,l}\Phi_{\alpha ,l}
^{[2,1,0,1]}\omega^{1-l}_{1-l}=[\sum\limits_{r=1}^{\lambda-1}\kappa_r
K^r (1+x^r )(x^2 -1)-2]\omega^{1}_{1}, \eeq with $x$ is given by
(17). The corresponding $C_{\lambda}$-extended Casimir operator is
\beq C(\lambda)=(\omega^{1}_{1})^{2}-\frac{1}{2}\{\omega^{2}_{1},
\omega^{0}_{1}\},\eeq evidently the $C_{\lambda}$-extended Casimir
don't commute with all generators (53-55), for example \beq
\begin{array}{ll}
[C(\lambda),\omega^{1}_{1}]&=-\frac{1}{2}[\{\omega^{2}_{1},
\omega^{0}_{1}\}, \omega^{1}_{1}]\\ \\
&=\frac{1}{2}\Big [ \sum\limits_{r=1}^{\lambda-1} \kappa_r K^r
(x^{3r}-(1+x^r )x
-(x^r +x^{2r})x+1)(x-1)\Big ] \omega^{2}_{2}\\ \\
&-\Big [ \sum\limits_{r=1}^{\lambda-1} \kappa_r K^r (x^r
+x^{2r})x-1\Big ] \Big [
1+\frac{1}{2}\sum\limits_{r=1}^{\lambda-1} \kappa_r K^r(1+x^r
)\Big ](x-1) \omega^{1}_{1}
\end{array}\eeq

\section {Conclusion and remarks}
\hspace{.3in}In the present paper, we have found the new deformed
$w_{\infty}$-symmetry  by adopting the approach for undeformed
case where the elements of the $w_{\infty}$-algebras
$\omega^{s}_{m}$ may be expressed via the pair $a^{\dagger}$ and
$a$ of quantum oscillator. We have shown that in the
Calogero-Vasiliev case $\lambda =2$, only in the case associated
with odd and even indices the obtained symmetries are changed.
Finally, note that in the same way one can construct the
$C_{\lambda}$-extended oscillator algebras of the supersymmetric
extension of $w_{\infty}$-algebra by using the super-realization
of $C_{\lambda}$-extended oscillator algebras.
\section*{Acknowledgements}
\hspace{.3in}J. Douari acknowledges the hospitality of the IHES,
France. E. H. El Kinani is grateful to the Abdus-Salam ICPT,
Trieste, Italy for its hospitality.

\end{document}